\documentclass[letterpaper, 10 pt, conference]{ieeeconf}
\IEEEoverridecommandlockouts
\def\BibTeX{{\rm B\kern-.05em{\sc i\kern-.025em b}\kern-.08em
    T\kern-.1667em\lower.7ex\hbox{E}\kern-.125emX}}

\usepackage{graphicx} 
\usepackage[none]{hyphenat}
\usepackage{amsmath}
\usepackage{amssymb}
\usepackage{mathrsfs,bm}
\usepackage{cite}
\usepackage{graphicx}
\usepackage{amsfonts}
\usepackage{amstext}
\usepackage{float}
\usepackage{cleveref}
\usepackage{nomencl}
\newtheorem{theorem}{Theorem}

\newtheorem{assumption}{Assumption}
\newtheorem{definition}{Definition}
\newtheorem{lemma}{Lemma}
\usepackage{color}

\begin{document}

\title{ Differential Privacy in Nonlinear Dynamical Systems with Tracking Performance Guarantees \\
\author{Dhrubajit Chowdhury, Raman Goyal,  and Shantanu Rane
\thanks{D. Chowdhury, R. Goyal, and S. Rane are with Palo Alto Research Center - Part of SRI International, Palo Alto, CA, USA. 
{ \tt\small \{dhruba.chowdhury, raman.goyal, shantanu.rane\}@sri.com }}}
}

\maketitle

\begin{abstract}
We introduce a novel approach to make the tracking error of a class of nonlinear systems differentially private in addition to guaranteeing the tracking error performance. We use funnel control to make the tracking error evolve within a performance funnel that is pre-specified by the user. We make the performance funnel differentially private by adding a bounded continuous noise generated from an Ornstein-Uhlenbeck-type process. Since the funnel controller is a function of the performance funnel, the noise adds randomized perturbation to the control input. We show that, as a consequence of the differential privacy of the performance funnel, the tracking error is also differentially private. As a result, the tracking error is bounded by the noisy funnel boundary while maintaining privacy. We show a simulation result to demonstrate the framework.

\end{abstract}
\section{Introduction}
The use of Cyber-Physical Systems (CPS) in our daily lives has been rapid recently due to the advancement in sensing and computational power. CPS finds widespread applications in various domains, including intelligent transportation systems, smart homes, and even the development of smart cities. However, these systems heavily rely on user-generated data to make informed decisions, thereby increasing the vulnerability of sensitive user information to potential exposure.
To address the concern of protecting sensitive user data several privacy-preserving frameworks, namely, differential privacy, information-theoretic privacy, and privacy based on secure multiparty computation have been developed. See the survey paper \cite{hassan2019differential} for a comprehensive overview of privacy algorithms.
\subsection{Background on Differential Privacy}
Differential privacy is a statistical notion of privacy that masks sensitive data using a mechanism that makes the output of the mechanism approximately unchanged if data belonging to any single user in the database is modified \cite{han2018privacy}. One of the main advantages of differential privacy is its protection from post-processing and it is not weakened even if an adversary knows the privacy mechanism used \cite{dwork2014algorithmic,le2013differentially}. Differential privacy is introduced using the “input perturbation” approach, which essentially means that noise is added to the system in either the input or output. However, adding noise to the system leads to a degradation in system performance both in static and dynamic cases \cite{le2013differentially}. In dynamical systems differential privacy makes the state trajectory of the system approximately indistinguishable \cite{le2013differentially} from other nearby state trajectories which the system could have produced.

Differential privacy was initially intended \cite{dwork2014algorithmic,dwork2008differential} for protecting the information of individuals within static databases. It has since evolved to handle the privacy issues in control \cite{han2018privacy} and dynamical systems. Recent work on privacy in linear dynamical systems includes dynamic filters \cite{le2013differentially}, differentially private linear quadratic (LQ) control \cite{yazdani2022differentially}, multi-agent formation control \cite{hawkins2022differentially}, and privacy-preserving consensus \cite{mo2016privacy,huang2012differentially}. However, unlike linear systems, the research in differential privacy in nonlinear systems is limited. In \cite{kawano2018differential,kawano2020design}, differential privacy was shown for incrementally input-to-state stable nonlinear systems without any performance guarantees.
\subsection{Main Contribution}
\begin{figure}[h!]
    \centering
    \includegraphics[width=\linewidth]{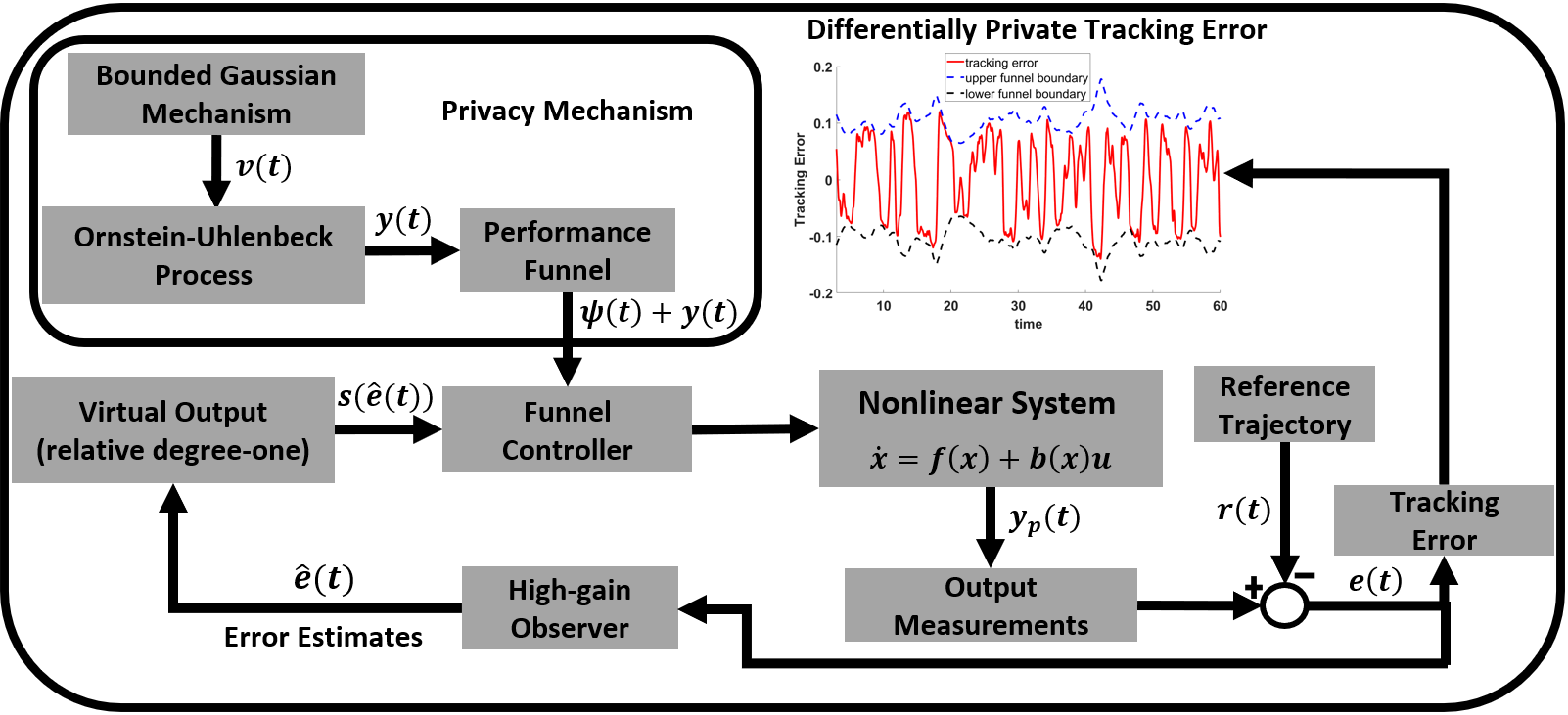}
    \caption{Design architecture for making the tracking error differentially private by adding privacy noise to the performance funnel.}
    \label{Fig:Architecture}
\end{figure}

The main contributions of this paper are:
\begin{itemize}
	\item  We develop a new framework for making the tracking error of nonlinear systems differentially private using a funnel controller  \cite{ilchmann2002tracking}. We add the privacy noise to the performance funnel (refer Fig. \ref{Fig:Architecture}) to make it differentially private. Since the controller is an explicit function of the performance funnel \cite{chowdhury2019funnel}, we indirectly add privacy noise to the control input of the system.
	\item The privacy noise which is added to the performance funnel to make it differentially private is continuous and bounded. The performance funnel \cite{chowdhury2019funnel} belongs to a class of weak differentiable functions, therefore we cannot add privacy noise directly to the performance funnel as the noise is discontinuous in nature. The noise is filtered through an Ornstein-Uhlenebck type process which makes it continuously differentiable and then it is added to the performance funnel.
	\item We use the funnel control algorithm developed in \cite{chowdhury2019funnel} for controlling the transient behavior of the tracking error for nonlinear systems with arbitrary relative degrees using high-gain observers. Since the tracking error evolves within the performance funnel we show using adjacency relationship that the tracking error becomes differentially private.
\end{itemize}

\section{Tracking Using Funnel Control}
We use the funnel control algorithm developed in \cite{chowdhury2019funnel} for controlling the transient behavior of the tracking error for nonlinear systems with arbitrary relative degrees using high-gain observers. The funnel controller algorithm in \cite{chowdhury2019funnel} is motivated by \cite{ilchmann2002tracking} where the tracking error evolves within a performance funnel, but it differs from typical funnel controllers in the literature where the controller does not require tuning of certain design parameters. By using the state measurements we synthesize a virtual output which makes the system relative degree-one with respect to the virtual output. The virtual output design ensures that by maintaining it in the performance funnel we can maintain the tracking error evolve within the funnel. Assuming that the state derivatives are not available for feedback we use a high-gain observer to estimate the derivatives and construct the virtual output.
\subsection{Funnel Control}
\begin{figure}[h]
\centering
 \includegraphics[width=0.9\linewidth]{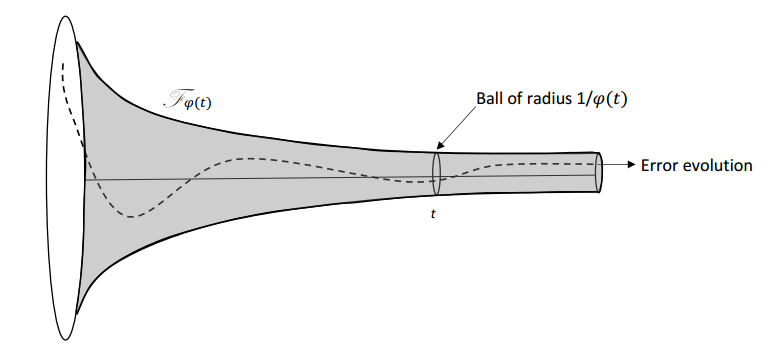}
\caption{Performance Funnel $\mathcal{F}_{\varphi}$ }
\label{Fig:Funnel_Pic}
\end{figure}

The concept underlying funnel control revolves around the use of a performance funnel to regulate the transient behavior of tracking errors. This is achieved by exploiting the inherent high-gain characteristic of the system. When the error approaches the funnel's boundary, the gain is elevated, preventing the error from reaching the boundary. Fig. \ref{Fig:Funnel_Pic} provides a visual representation of a performance funnel $\mathcal{F}_{\varphi}$ and the error evolution within it.
 Let $\varphi$ be a function of the following class:
 $\bar{\Phi}\colon=\{\varphi\in W^{1,\infty}(\mathbb{R}_{\geq 0},\mathbb{R}_{+})\hspace{0.1 cm}| \hspace{0.1 cm} \forall \ \tau\geq0 :\hspace{0.1 cm} \varphi(\tau)>0 \hspace{0.1 cm} \text{and} \ \lim_{\tau \to \infty} \text{inf}\hspace{0.1 cm} \varphi(\tau) >0 \}$
 where $W^{1,\infty}(\mathbb{R}_{\geq 0},\mathbb{R}_{+})$ represents the class of weakly differentiable functions. The performance funnel is defined as :
\begin{equation}
\mathcal{F}_{\varphi}:= \{(t,e)\in\mathbb{R}_{\geq 0}\times\mathbb{R}\hspace{0.1 cm}|\hspace{0.1 cm}\varphi(t) |e|<1\}
\end{equation}

The reciprocal of the function $\varphi(t)$ determines the funnel boundary and the error $e(t)$ evolves within the funnel $\mathcal{F}_{\varphi}$. In this paper, we assume that the funnel is finite i.e., $\varphi(0) > 0$ and we define the funnel boundary as $\psi(t)=1/\varphi(t)$.

\subsection{Tracking Problem Definition}
We consider the tracking problem for a single-input-single-output system, which is defined globally in the normal form \cite{khalil2002nonlinear} :
\begin{subequations}
\begin{align}
\begin{split}
\dot{\xi}_i&=\xi_{i+1},\hspace{0.1 cm} 1\leq i \leq \rho -1
\end{split}\\
\begin{split}
\dot{\xi}_{\rho}&=a(t,\xi)+b(\xi)u
\end{split}\\
\begin{split}
y_p&=\xi_1
\end{split}
\end{align}
\label{eq:plant}
\end{subequations}
where $\xi=\text{col}(\xi_1,\xi_2,\ldots,\xi_{\rho}) \in R^{\rho}$, $u \in R$ and  $y_p \in R$. The function $b$ is locally Lipschitz; $a$ is locally Lipschitz in $\xi$ and piecewise continuous, bounded in $t$. We can also extend this class of systems to the special normal form see \cite[Section 9.1]{isidori1985nonlinear}.
\begin{assumption}
$b(\xi)$ is known and satisfies
$$b(\xi)\geq b_{0}>0,\hspace{1 cm} \forall \ \xi \in R^{\rho}$$
\end{assumption}
\begin{assumption}
The reference signal  $r(t)$ and its derivatives up to $r^{(\rho)}(t)$ are bounded for all $t \geq 0$ and the $\rho$th  derivative  $r^{(\rho)}(t)$ is a piecewise continuous function of $t$.
\end{assumption}
We define $\mathcal{R}=\text{col}(r,r^{(1)},\ldots,r^{(\rho-1)})$ and assume that $r(t)$ is available for control.

Consider the following change of variables
$$\omega_1=\xi_1-r, \omega_2=\varrho(\xi_2-r^{(1)}), \ldots,\omega_{\rho}=\varrho^{\rho-1}(\xi_{\rho}-r^{(\rho-1)}) $$
where  $\varrho>0$ is a scaling variable. The change of variables in matrix form is defined by
$$\xi=L^{-1}(\varrho)\omega_f+\mathcal{R}$$
where
  $$L(\varrho)=\begin{bmatrix}
1 &  0  & 0 & \ldots  & 0\\
0  & \varrho & 0 & \ldots & 0\\
\vdots & \vdots & \ddots &\ddots &\vdots\\
0  & 0& \ldots  & \varrho^{\rho-2}  & 0\\
0  & 0  & \ldots &  \ldots & \varrho^{\rho-1}\\
\end{bmatrix}$$
and $\omega_f=\text{col}(\omega_1,\ldots,\omega_{\rho})$.
The change of variables transforms system (\ref{eq:plant}) into:
\begin{subequations}
\begin{align}
\begin{split}
\varrho\dot{\omega}_i&=\omega_{i+1}, \ \ 1\leq i \leq \rho -1
\end{split}\\
\begin{split}
\varrho\dot{\omega}_{\rho}&=\varrho^{\rho}\{a(t,\xi)+b(\xi)u-r^{(\rho)}(t)\}
\end{split}\\
\begin{split}
e&=\omega_1
\end{split}
\end{align}
\label{eq:sys_transform}
\end{subequations}
where $e=y_p-r$ is the tracking error and \linebreak
$\xi=\text{col}\left(\omega_1+r, \dfrac{\omega_2}{\varrho}+r^{(1)},\ldots, \dfrac{\omega_{\rho}}{\varrho^{\rho-1}}+r^{(\rho-1)}\right).$
\begin{assumption}
There exists a known continuous function $g_1(\cdot)$ such that
		$$|\varrho^{\rho-1}a(t,L^{-1}(\varrho)\omega_f+\mathcal{R})| \leq g_1(||\omega_f||)$$ 
		for $\varrho \in (0,\varrho_1)$, for some $\varrho_1 > 0$ and for all $t\geq0$.
\end{assumption}
Assumption 3 is satisfied when $a$ is globally Lipschitz in $\xi$. It is also satisfied if 
$$|a|\leq g_2(|\xi_1|)+b_1|\xi_2|^{\rho-1}+b_2|\xi_3|^{\frac{\rho-1}{2}}+\ldots+b_{\rho-1}|\xi_{\rho}|$$
where $g_2$ is locally Lipschitz in $\xi_1$ and $b_i$ for $i=1,\ldots, \rho-1$ are positive constants.
\subsection{Funnel Control by Synthesizing Virtual Output }
Funnel control was used to evolve the tracking error $e$ inside the performance funnel using the virtual output for known relative degree systems in \cite{chowdhury2019funnel}, \cite{chowdhury2017funnel}. Here we present an abridged version of the main idea.
\subsubsection{State Feedback Funnel Controller Design}
We synthesize the virtual output from the system states as
\begin{equation}
s=\omega_1+k_{2}\omega_{2}+\ldots+ k_{\rho}\omega_{\rho}
\label{eq:s}
\end{equation}
where $k_2$ to $k_{\rho}$ are positive constants to be chosen. The relative degree of the system (\ref{eq:sys_transform}) with respect to $s$ is one. By using feedback control we introduce a two-time scale structure to make $\omega_1,\omega_2,\ldots,\omega_{\rho-1}$ fast while making $s$ slow. By choosing $u$ as
\begin{equation}
u=\frac{1}{\varrho^{\rho}k_{\rho}b(\xi)}[-\omega_2-k_{2}\omega_{3}-\ldots-k_{\rho-1}\omega_{\rho}+\varrho v_f]
\label{eq:statefeedback1}
\end{equation}
where $v_f$ is an auxiliary input. The singularly perturbed system is given by
\begin{subequations}
\begin{align}
\begin{split}
\varrho\dot{\omega}&=F\omega+Hs
\end{split}\\
\begin{split}
\dot{s}&=v_f+\varrho^{\rho-1}k_{\rho}\{a(t,\xi)-r^{(\rho)}(t)\}
\end{split}
\end{align}
\label{eq:singularly_perturbed}
\end{subequations}
$$
F=\begin{bmatrix}
0 &  1  & \ldots & \ldots & 0\\
0  & 0 & 1 & \ldots & 0\\
\vdots & &  & &\vdots\\
0  & \ldots  & \ldots & 0 & 1\\
-\frac{1}{k_{\rho}}  & \ldots  & \ldots &  -\frac{k_{\rho-2}}{k_{\rho}} & -\frac{k_{\rho-1}}{k_{\rho}}\\
\end{bmatrix},  H=\begin{bmatrix}
0 \\
0 \\ 
\vdots \\
 0 \\
 \frac{1}{k_{\rho}}
\end{bmatrix}
$$
where $\omega=\text{col}(\omega_1,\ldots,\omega_{\rho-1})$, $F \in R^{\rho-1 \times \rho-1}$, $H \in R^{\rho-1 \times 1}$. The gains $k_2,k_3,\ldots,k_{\rho}$ are chosen such that the matrix $F$ is Hurwitz, which is always possible. In (\ref{eq:singularly_perturbed}), $\omega$ is the fast variable and $s$ is the slow variable. 
\begin{theorem}
\textit{Consider the closed-loop system (\ref{eq:singularly_perturbed}) obtained using the state feedback controller (\ref{eq:statefeedback1}). Let $k_2$ to $k_{\rho}$ be chosen such that the matrix $F$ is Hurwitz. Suppose Assumptions 1-3 are satisfied. Let $\varphi(t) \in \bar{\Phi}$ and suppose the initial states satisfy $(\omega(0),s(0)) \in \Lambda_0$, where $\Lambda_0$ is a compact set. Then the funnel controller,
 \begin{equation}
 v_f=-\frac{1}{\psi(t)-|s|}s
 \label{eq:statefeedback2}
 \end{equation}
  yields a closed-loop system such that the solution is bounded for all $t\geq0$ and there exists $\varrho^*>0$ such that for each $\varrho \in (0,\varrho^{*}]$, there exists $\kappa_s^*>0$ such that}
  \begin{equation}
  |e(t)| \leq \psi(t)-\kappa_s^*, \hspace{1 cm}\forall \  t\geq0
  \label{eq:e_ineq_state}
  \end{equation}
  
\end{theorem} 
 \textbf{Proof:} See \cite{chowdhury2019funnel}
\subsubsection{Output Feedback Funnel Controller Design}
The virtual output in the previous section is constructed from the knowledge of all the state variables. In this section we estimate $\hat{\mathcal{E}}=\text{col}(\hat{e}_1,\hat{e}_2,\ldots,\hat{e}_{\rho})$ by only measuring $e$ using a high-gain observer:
\begin{subequations}
\begin{align}
\begin{split}
\dot{\hat{e}}_{i}&=\hat{e}_{i+1}+\frac{\gamma_{i}}{\varsigma^i}(e-\hat{e}_1),\ \ 1\leq i \leq \rho-1
\end{split}\\
\begin{split}
\dot{\hat{e}}_{\rho}&=a_0(\hat{\xi})+b(\hat{\xi})\hat{u}_s+\frac{\gamma_{\rho}}{\varsigma^{\rho}}(e-\hat{e}_1)
\end{split}
\end{align}
\label{eq:hgo}
\end{subequations}
where $\varsigma$ is a small positive constant and $\gamma_1,\gamma_2,\ldots,\gamma_{\rho}$ are chosen such that the polynomial,
\begin{equation}
t^{\rho}+\gamma_{1}t^{\rho-1}+\ldots+\gamma_{\rho-1}t+\gamma_{\rho}
\label{eq:hurwitz}
\end{equation}
is Hurwitz, $a_0(\xi)$ serves as a nominal model for $a(t,\xi)$ and $\hat{\xi}_i=\hat{e}_i+r^{(i-1)}$. From the estimates $\hat{e}_1$ to $\hat{e}_{\rho}$ we have $\hat{\omega}_1=\hat{e}_1$, $\hat{\omega}_2=\varrho\hat{e}_2$, \ldots, $\hat{\omega}_{\rho}=\varrho^{\rho-1}\hat{e}_{\rho}$ and an estimate of the virtual output is given by 
\begin{equation}
\hat{s}=\hat{\omega}_{1}+k_{2}\hat{\omega}_{2}+k_{3}\hat{\omega}_{3}+\ldots+k_{\rho}\hat{\omega}_{\rho }    
\label{eq:voutput}
\end{equation}
 We saturate the estimates $\hat{\omega}_1,\ldots,\hat{\omega}_{\rho}$ outside the compact set $\Lambda_f$ to overcome the peaking phenomenon of the observer \cite{khalil2017high}. The estimates are saturated as $\hat{\omega}_{is}=\bar{M}_i\text{sat}\left(\frac{\hat{\omega}_{i}}{\bar{M}_i}\right)$ where sat is the saturation function \cite{khalil2017high}. The estimate of the virtual output is saturated as,
$$\hat{s}_s=\hat{\omega}_{1s}+k_{2}\hat{\omega}_{2s}+k_{3}\hat{\omega}_{3s}+\ldots+k_{\rho}\hat{\omega}_{\rho s}$$
The funnel controller gain is given by $\hat{k}(t)=\dfrac{1}{\psi(t)-|\hat{s}_s|}$ and the funnel gain is saturated as $\hat{k}_s=\bar{M}_k\text{sat}\left(\dfrac{\hat{k}}{\bar{M}_k}\right)$.

See \cite{chowdhury2019funnel} on how to select the saturation levels $\bar{M}_1, \ldots, \bar{M}_{\rho}$ and $\bar{M}_k$. Using the estimates $(\hat{\omega}_1,\hat{\omega}_2,\ldots,\hat{\omega}_{\rho})$, the output feedback control is given by,
\begin{subequations}
\begin{align}
\begin{split}
\hat{u}_s&=\frac{1}{\varrho^{\rho}k_{\rho}b(\hat{\xi})}[-\hat{\omega}_{2s}-\ldots-k_{\rho-1}\hat{\omega}_{\rho s}+\varrho \hat{v}_{fs}]
\end{split}\\
\begin{split}
\hat{v}_{fs}&=-\hat{k}_s\hat{s}_s
\end{split}
\end{align}
\label{eq:outputfeedback}
\end{subequations}
 \begin{theorem}\textit{
Consider the plant (\ref{eq:plant}), the observer (\ref{eq:hgo}), and the output feedback controller (\ref{eq:outputfeedback}). Suppose all the assumptions of Theorem 1 are satisfied, $\gamma_1$ to $\gamma_{\rho}$ are chosen such that the polynomial (\ref{eq:hurwitz}) is Hurwitz and $\hat{\mathcal{E}}(0) \in Y$ where $Y$ is a compact subset of $R^{\rho}$. Then there exists $\varrho^{**}>0$ and for each $\varrho \in (0,\varrho^{**})$, there is $\varsigma^*=\varsigma^*(\varrho)>0$, such that for each $\varrho \in (0,\varrho^{**})$ and $\varsigma \in (0,\varsigma^ *(\varrho))$ there exists $\kappa_o^*>0$ such that
\begin{equation}
|e(t)| \leq \psi(t)-\kappa_o^*, \hspace{1 cm}\forall \  t\geq0
\label{eq;e_ineq_out}
\end{equation}
and the states $(\omega(t),s(t),\hat{\mathcal{E}})$ of the closed loop system are bounded for all $t \geq 0$}. 
 \end{theorem}
\textbf{Proof:} See \cite{chowdhury2019funnel}

\section{Differential Privacy Problem Setup}
\subsection{Adjacency Relation}
We consider funnel boundary trajectories of the form $\psi(t)=(\psi(t_1), \psi(t_2), \ldots)$, where $\psi(t) \in \mathbb{R}$ and $0<\psi(t)<\infty$ for all $t \geq 0$. We denote the set of all such sequences by $\psi \in {\ell}_1$. We will define our adjacency relation over  ${\ell}_1$. 
\begin{definition}
    (Adjacency for funnel boundary): We define the dataset $\Psi$ and $\Psi^{\prime}$ where each element in the set is the tuple $\{-\psi(t_i),\psi(t_i)\}_{i=1}^n$ and $\{-\psi^{\prime}(t_i),\psi^{\prime}(t_i)\}_{i=1}^n$. The two trajectories $\psi, \psi^{\prime} \in {\ell}_1$ are adjacent if
    
    \begin{equation}
        |\psi(t_i)-\psi^{\prime}(t_i)| \leq \delta \psi, \ \  \forall \ 1 \leq i \leq n
          \label{eq:adj}
    \end{equation}
    where $\delta \psi > 0$ is the adjacency parameter. The value of $n$ is chosen based on some finite time $T_f>0$ such that for all $t_i \in[0,T_f]$ the user requires differential privacy. We will write $\operatorname{Adj}_{\delta \psi}(\psi ,\psi^{'})=1$ if $\psi$ and $\psi^{'}$ are adjacent which implies (\ref{eq:adj}) holds, and $\operatorname{Adj}_{\delta \psi}(\psi, \psi^{'})=0$, otherwise. The constant $\delta \psi$ is chosen based on the privacy requirement as the adjacency relation implies that a particular funnel boundary can be made approximately indistinguishable within distance $\delta \psi$ from all other funnel boundaries. 
    \end{definition}
\begin{definition}
We define a query as 
 \begin{equation}
  Q(d)= \psi   
  \label{eq:query}
 \end{equation}
 where $d \in \Psi$. The query implies that each time the query is called one funnel boundary is selected from the database.
\end{definition}
\begin{definition}
(Query Sensitivity): The sensitivity of the query $Q$ is given by
\begin{equation}
\nonumber  \Delta Q := \sup_{d,d^{'}|\operatorname{Adj}_{\delta \psi} (d,d^{'})=1} |Q(d)-Q(d^{'})|
\label{eq:query_sensitivity}
\end{equation}


The sensitivity captures the largest magnitude by which the output of the query can change across two adjacent databases.
\end{definition}
Next, we define differential privacy for dynamic systems (see \cite{le2013differentially} for a formal construction).
\begin{definition}
(Differential privacy for funnel boundary/tracking error): Let $\epsilon>0$ and $\delta \in(0,1 / 2)$ be given. A mechanism $\mathcal{M}$ is $\left(\epsilon, \delta\right)$-differentially private if, for all adjacent $\psi, \psi^{\prime} \in \ell_1$ or $e, e^{\prime} \in \ell_1$ , we have:
$$
    \nonumber \mathbb{P}\left[\mathcal{M}\left(\psi\right) \in S\right] \leq \exp^{\epsilon} \mathbb{P}\left[\mathcal{M}\left(\psi^{\prime}\right) \in S\right]+\delta \text { for all } S \in \mathbb{R} .
$$
$$
    \nonumber \mathbb{P}\left[\mathcal{M}\left(e\right) \in S\right] \leq \exp{^{\epsilon}} \mathbb{P}\left[\mathcal{M}\left(e^{\prime}\right) \in S\right]+\delta \text { for all } S \in \mathbb{R} .
$$
\end{definition}



\subsection{Univariate Bounded Gaussian Noise}
A mechanism will add noise to the funnel boundary to make the funnel boundary differentially private. However, adding arbitrary noise to the funnel boundary can violate the assumptions of the funnel boundary. For example, if $\psi(t)+v(t)<0$ for any $t\geq 0$, the funnel controller will fail to work. Therefore, in this section, we generate a bounded Gaussian noise. The bounded domain is given by  $\mathcal{D}=[\alpha,\beta] \subset \mathbb{R}$, where $\alpha=-c_1\psi_{\min}$, $\psi_{\min}=\underset{t\geq 0}{\inf} \ \psi (t)$, $0<c_1<1$, and $\beta=-\alpha$ is chosen to make the probability density function symmetric. In general, $\beta$ can be chosen as $\beta=c_2\psi_{\max}$, where $c_2 \geq 1$ and $\psi_{\max}=\underset{t\geq 0}{\sup} \ \psi (t)$.
\begin{definition}
(Univariate bounded Gaussian noise): Given $\mathcal{D}=[\alpha,\beta]$ where $(\alpha<\beta)$, both finite is a constrained domain. Then the probability density of the univariate Gaussian noise is given by
\begin{equation}
  p_{B}(v) =
    \begin{cases}
         \dfrac{1}{\sigma} \dfrac{\phi\left(\dfrac{v}{\sigma}\right)}{\Phi(\beta^{'})-\Phi(\alpha^{'})} \ \ \text{if} \ v \in \mathcal{D}, \\
      0 \hspace{25mm} \text{otherwise},
    \end{cases}  
    \label{eq:bounded_Gauss}
\end{equation}
where the original Gaussian distribution is zero mean and $\sigma$ variance, $\beta^{'}= \dfrac{\beta}{\sigma}$, $\alpha^{'}= \dfrac{\alpha}{\sigma}$ and 
$$\phi(v) = \dfrac{1}{\sqrt{2\pi}} \exp{\left(-\dfrac{1}{2} v^2\right)}, \ \Phi(v)=\dfrac{1}{2}\left(1+\text{erf}(v/\sqrt{2})\right)$$
\end{definition}\vspace{5mm}

We cannot sample noise from the distribution (\ref{eq:bounded_Gauss}) and add it to the funnel boundary as it will make the performance funnel discontinuous. Therefore, we require the noise to be filtered before adding it to the funnel boundary which is discussed in the next section.

\section{Ornstein-Uhlenbeck Type Process}
The bounded Gaussian noise is passed through a linear filter to produce a continuous noise which is added to the funnel boundary. We model this operation as an Ornstein–Uhlenbeck (OU) type process which is a stationary process \cite{risken1985fokker}. It has the property that over time, the process tends to drift towards its mean function: such a process is called mean-reverting. We define the system as
\begin{subequations}
\begin{align}
\begin{split}
\theta \dfrac{dy}{dt}= -y(t)+w(t)
\end{split}\\
\begin{split}
\dfrac{dw}{dt}= -\vartheta w(t)+v(t)
\end{split}
\end{align}
\label{eq:OU}
\end{subequations}
where $0<\theta<<1$, $\vartheta>0$. 
The system (\ref{eq:OU}) is represented by a singularly perturbed system where $y$ is the fast variable and $w$ is the slow variable. 
\subsection{Boundedness \& Solutions of the Process}
The solution of the decoupled equation (\ref{eq:OU}b) is given by
\begin{equation}
 w(t)=w_0e^{-\vartheta t}+  \int_0^t e^{-\vartheta(t-s)}v(s)ds
\end{equation}
We choose the initial condition of the process as $w_0=0$ for simplicity. The solution can then be defined as 
$$w(t)= \int_0^t e^{-\vartheta(t-s)}v(s)ds$$
The solution is an integral of a deterministic function with respect to a bounded Gaussian noise. From (\ref{eq:bounded_Gauss}) we have 
$\alpha \leq v(t) \leq \beta$ which implies
\begin{equation}
\dfrac{\alpha}{\vartheta}(1-e^{-\vartheta t}) \leq     w(t) \leq \dfrac{\beta}{\vartheta }(1-e^{-\vartheta t}), \ \forall \ t\geq 0
\label{eq:w bounds}
\end{equation}





\begin{theorem}
\textit{Let OU type process be defined by (\ref{eq:OU}). Let $v(t)$ be the noise generated from the truncated Gaussian probability distribution (\ref{eq:bounded_Gauss}). Then, there exists a time $T(\theta)$ such that for all $t \geq T(\theta)$ where $\underset{\theta \rightarrow 0 }{\lim} \ T(\theta)=0$ }
\begin{equation}
    y(t)= w(t) + O(\theta) , \ \ t\geq T(\theta)
    \label{eq:approx}
\end{equation}
\end{theorem}
\textbf{Proof:} The quasi-steady state of (\ref{eq:OU}a) is obtained by setting $\theta =0$, from which we have
$$y(t)=w(t)$$
Next we define the variable $z(t)=y(t)-w(t)$ and by taking its derivative we have
\begin{equation}
    \theta \dot{z}=-z- \theta[-\vartheta w(t)+v(t)]
\end{equation}
By defining a Lyapunov function $V_{z}=(1/2)z^2$, and taking its derivative along (\ref{eq:OU}), we have
$$\theta\dot{V}_z \leq -z^2+\theta \bar{\Delta}|z|$$
where $|-\vartheta w(t)+v(t)| \leq \bar{\Delta}$, where the right-hand side is bounded since (\ref{eq:OU}b) is Bounded-Input-Bounded-Output (BIBO) stable system. From which we have 
$$\dot{V}_z \leq -\dfrac{z^2}{2\theta}, \ \ \forall \ |z| \geq 2\theta \bar{\Delta}$$
It can be shown \cite{khalil2017high}, that there exists a time $T(\theta)>0$, where $\lim_{\theta \rightarrow 0} T(\theta)=0$ such that
$$|z(t)| \leq 2\theta \bar{\Delta}, \ \ \forall \ t \geq T(\theta)$$
From which we can conclude that (\ref{eq:approx}) follows.
\subsection{Probability Density Function of the OU type Process}
We first discuss the probability density functions (pdf) of $w(t)$, and $y(t)$ when the input noise is sampled from a Gaussian distribution. There will be a steady-state pdf for the variables $w(t)$, and $y(t)$ as the process (\ref{eq:OU}b) is stationary \cite{risken1985fokker} since $\vartheta>0$. We define the following lemma. 

\begin{lemma}
\textit{Let $v_G$ be the input to the OU type process (\ref{eq:OU}), where $v_G$ is sampled from a Gaussian distribution $v_G \sim  \mathcal{N}(0,\sigma_G^2)$ with pdf $p_G(v_G)=\dfrac{1}{\sigma_G} \phi\left(\dfrac{v_G}{\sigma_G}\right)$. Then, the pdf of $y$ up to an order of $O(\theta)$ for all $t \geq T_{ss}$ is given by 
\begin{equation}
    p_{ss}(y)=\dfrac{1}{\sigma_G^{\prime}} \phi\left(\dfrac{v}{\sigma_G^{\prime}}\right)
    \label{eq:fokker-planck}
\end{equation}
where $p_{ss}$ is the steady-state pdf of y, $\sigma_G^{\prime}=\sigma_G/(\sqrt{2\vartheta})$ and the mechanism $\mathcal{M} = Q(d)  + y(t)$, where $d \in \Psi$, makes the funnel boundary $(\epsilon, \delta)$-differentially private with
$\sigma_G^{\prime} \geq \Delta Q \delta \psi\kappa\left(\delta_G, \epsilon_G\right)$, where   $\kappa\left(\delta_G, \epsilon_G\right)=\frac{1}{2 \epsilon_G}\left(K_{\delta_G}+\sqrt{K_{\delta_G}^2+2 \epsilon_G}\right)$, with $K_{\delta}:=\mathcal{Q}^{-1}\left(\delta_G\right)$, $\mathcal{Q}$ representing the Gaussian tail integral.}
\end{lemma}
\textbf{Proof:} When $v_G$ is sampled from a Gaussian distribution we can rewrite (\ref{eq:OU}b) as
$$dw= \vartheta (\mu -w(t))dt+\sigma dq$$
where $q$ is a Wiener process \cite{risken1985fokker}. We can use Ito's integral \cite{risken1985fokker} to show that conditional expectation and variance for $w_0=0$ are
$$E[w(t)] = E\left[ \int_0^t e^{-\vartheta(t-s)}dq(s)\right] =\mu(1-e^{-\vartheta t}) =0$$
since $\mu=0$, and 
$$ \text{Var}[w(t)] = E\left[ \left(\sigma_G^{\prime}\int_0^t e^{-\vartheta(t-s)}dq(s) \right)^2 \right]  ={\sigma_{G}^{\prime}}^2(1-e^{-2\vartheta t}).$$ It can be shown that the probability distribution using the Fokker-Planck representation \cite{risken1985fokker} is given by
$$p(w,t)=\dfrac{1}{\sqrt{2\pi(1-e^{-2\vartheta t})}\sigma_G^{\prime}} \exp \left[\dfrac{ y^2}{{\sigma_G^2}^{\prime}  (1-e^{-2\vartheta t})}\right] $$
Then, for all $t\geq T_{ss}$, the steady-state probability distribution up to an error of $O(\theta)$ is given by (\ref{eq:fokker-planck}) using Theorem 3. Using the steady-state probability (\ref{eq:fokker-planck}), adjacency definition (\ref{eq:adj}) and query sensitivity (\ref{eq:query_sensitivity}), it can be shown \cite{dwork2008differential} that the mechanism $\mathcal{M} = Q(d)  + y(t)$, where $d \in \Psi$, makes the funnel boundary $(\epsilon, \delta)$-differentially private with
$\sigma_G^{\prime} \geq \Delta Q \delta \psi\kappa\left(\delta_G, \epsilon_G\right)$, where   $\kappa\left(\delta_G, \epsilon_G\right)=\frac{1}{2 \epsilon_G}\left(K_{\delta_G}+\sqrt{K_{\delta_G}^2+2 \epsilon_G}\right)$.

\textbf{Remark 1:} In our scenario, we cannot use Gaussian noise as the input to the OU type process as it will it might violate the funnel controller assumptions. Since we use truncated Gaussian noise as an input to the OU type process the pdf $p_{OU}(y)$ is not infinite support compared to $p_{ss}(y)$. But we conjecture that the shape of $p_{OU}(y)$ will be similar to the shape of $p_{ss}(y)$.

When the process (\ref{eq:OU}b) is driven by an input noise sampled from a truncated Gaussian probability distribution, the output $y(t)$ will have a  steady-state pdf. This follows since the process (\ref{eq:OU}b) is stationary. We denote the steady-state pdf of $y$ as $p_{OU}(y)$ for all $t \geq T_{ss}$.

\begin{theorem}
\textit{Let OU type process be defined by (\ref{eq:OU}). Let $v(t)$ be the noise generated from the truncated Gaussian probability distribution (\ref{eq:bounded_Gauss}). Then for all $t \geq T_{ss}$, 
\begin{itemize}
    \item $p_{OU}(y)$ is continuous
    \item  $p_{OU}(y)$ is bounded 
    \item $p_{OU}(y \geq \beta/\vartheta )=p_{OU}(y\leq \alpha/\vartheta)=0$ 
\end{itemize}}
\end{theorem}
\textbf{Proof:} We perform our analysis when the probability distribution $p(y)$ reach steady state for $t \geq T_{ss}$. The pdf $p_{OU}(y)$ is continuous since the random variable $y$ is continuous as it is the output of the OU type process (\ref{eq:OU}).

The boundedness of the pdf comes from the boundedness of the noise $v(t)$ which limits the range of $w(t)$ to (\ref{eq:w bounds}).

Finally, the upper and lower bounds of $w(t)$ are reached when $v(t)=\beta, \forall \ t \geq 0$ or $v(t)=\alpha, \forall \ t \geq 0$. The probability of sampling the bounds are as follows
$$P(w=\alpha)=\int_{\alpha}^{\alpha}p_B(v)dv=\int_{0}^{0}p_B(v)dv=0$$
$$P(w=\beta)=\int_{\beta}^{\beta}p_B(v)dv=\int_{0}^{0}p_B(v)dv=0$$
From which we can conclude that $p_{OU}(y=\beta/\vartheta )=p_{OU}(y=\alpha/\vartheta)=0$. For $t\geq T_{ss}$, the maximum and minimum value of $w(t)$ is $\beta/\vartheta$ and $\alpha/\vartheta$ from which we can conclude that $p_{OU}(y \geq \beta/\vartheta )=p_{OU}(y \leq \alpha/\vartheta)=0$.


\section{Differential Privacy of Funnel Boundary and Tracking Error }
\subsection{Funnel Boundary Differential Privacy}
\begin{theorem}
\textit{Let $v(t)$ be the noise generated from the pdf (\ref{eq:bounded_Gauss}) which is filtered through the OU type process (\ref{eq:OU}) with the output as $y(t)$ and pdf as $p_{OU}(y)$. Then, the mechanism  $\mathcal{M} = Q(d)  + y(t)$, where $d \in \Psi$, makes the funnel boundary $\left(\epsilon, \delta\right)$-differentially private with respect to $\operatorname{Adj}_{\delta \psi}$ in $\psi$ where}
\begin{equation}
    \epsilon \leq \epsilon_U, \  \ \delta \leq \delta_U
    \label{eq:inequality}
\end{equation}
\textit{for all} $t \in  [ \bar{T}, T_f]$ \textit{where} $\epsilon_U$, and $\delta_U$ are the upper bounds of $\epsilon$, and $\delta$, $\bar{T}=\max\{T(\theta),T_{ss}\}$ \textit{and} $T_f>\bar{T}>0$.
\end{theorem}

\begin{figure}[h]
 \includegraphics[width=\linewidth]{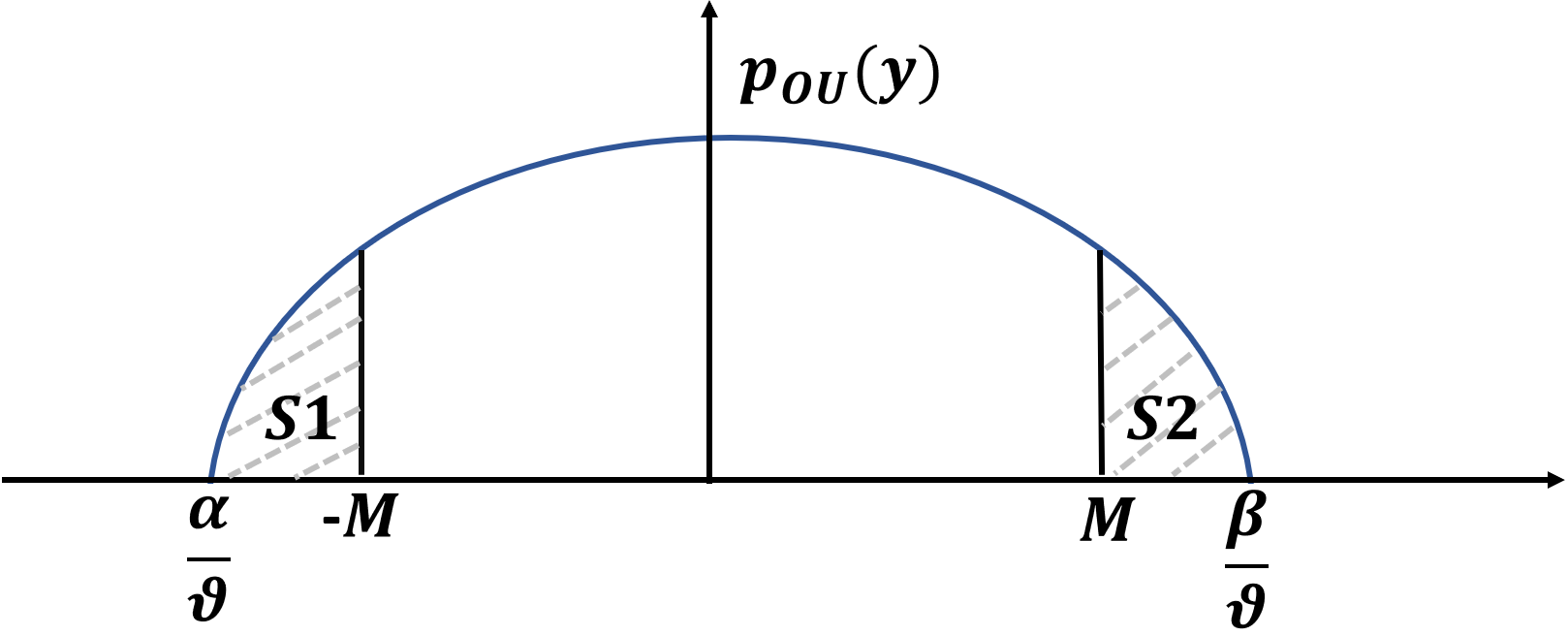}
\caption{ Probability density function of OU type process}
\label{Fig:pdf}
\end{figure}

\textbf{Proof:} We use the results in \cite{he2020differential} to find the upper bounds of $\epsilon$ and $\delta$. We perform our analysis when the pdf of $w(t)$ is in steady-state and (\ref{eq:approx}) holds. Therefore, our analysis is between the time period $t \in  [ \bar{T}, T_f]$ where $\bar{T}=\max\{T(\theta),T_{ss}\}$ and $T_f>\bar{T}>0$. For $t_i \in[ \bar{T}, T_f]$, we define the dataset $\bar{\Psi}$ and $\bar{\Psi}^{\prime}$ where each element in the set is the tuple $\{-\psi(t_i),\psi(t_i)\}_{i=n_1}^{n_2}$ as $\{-\psi^{\prime}(t_i),\psi^{\prime}(t_i)\}_{i=n_1}^{n_2}$ where $n_1$ and $n_2$ can be determined from $\bar{T}$, and $T_f$. During this time period we have
    \begin{equation}
        |\psi(t_i)-\psi^{\prime}(t_i)| \leq \delta \psi^{\prime}, \ \  \forall \ n_1 \leq i \leq n_2
          \label{eq:adj_new}
    \end{equation}
where $\delta \psi^{\prime}>0$ is the adjacency parameter in this time period. The steady-state pdf of $p_{OU}(y)$ is represented by Fig. \ref{Fig:pdf}. Following \cite[Theorem 3.6]{he2020differential}, we define $\Theta_0=[-M,M]$ and $\Theta_1=(-\infty,-M] \cup [M,\infty)$ such that
\begin{subequations}
\begin{align}
\begin{split}
\epsilon \leq \ln\left[ \underset{\delta \in [- \delta \psi^{\prime},  \delta \psi^{\prime}], y \in \Theta_0}{\sup}  \ \dfrac{p_{OU}(y-\delta )}{p_{OU}(y)} \right]
\end{split}\\
\begin{split}
 \delta \leq 2 \oint_{\Theta_1} p_{OU}(y)dy = S1 + S2
\end{split}
\end{align}
\label{eq:epsilon_delta}
\end{subequations}
Let $c_b=\underset{\delta \in [- \delta \psi^{\prime},  \delta \psi^{\prime}], y \in \Theta_0}{\sup}  \ \dfrac{p_{OU}(y-\delta )}{p_{OU}(y)}$. The supremum can be written as 
$$\dfrac{\underset{\delta \in [- \delta \psi^{\prime},  \delta \psi^{\prime}], y \in \Theta_0}{\max} \ p_{OU}(y-\delta )}{\underset{y \in \Theta_0}{\min}\ p_{OU}(y)} $$
The above holds because the pdf of $p_{OU}(y)$ is continuous and defined in a closed and compact interval. We consider two cases:\\
\textbf{Case I:} $ \delta \psi^{\prime}< M$: In this case $c_b$ will depend on  $ \delta \psi^{\prime}$ and $M$ as we have $c_b = \dfrac{p_{OU}(M- \delta \psi^{\prime})}{p_{OU}(M)}$\\
\textbf{Case II:} $ \delta \psi^{\prime}\geq M$: In this case $c_b$ will depend on $M$ as we have $c_b = \dfrac{p_{OU}(0)}{p_{OU}(M)}$\\
Taking $\epsilon_U =\ln (c_b)$, we arrive at the inequality. 

From Fig. \ref{Fig:pdf}, the term $\delta_U$ is bounded since the pdf $p_{OU}(y)$ is bounded and
$$2 \oint_{\Theta_1} p_{OU}(y)dy = S1 + S2$$

\subsection{Tracking Error Differential Privacy for State and Output feedback Funnel Controller}
\begin{theorem}
\textit{Let Theorem 1 and 5 hold and the state feedback funnel controller be defined by (\ref{eq:statefeedback1}) and (\ref{eq:statefeedback2}). Let the mechanism $\mathcal{M}:Q(d)+ y$ where $y$ is generated from the probability distribution (\ref{eq:bounded_Gauss}) and the OU type process (\ref{eq:OU}) makes the funnel boundary $(\epsilon,\delta)$ differentially private. Then, the tracking error $e$ obtained using the state feedback funnel controller is $(\epsilon,\delta)$ differentially private for all $t \in [\bar{T},T_f]$ where time $T_f>\bar{T}>0$.}
\end{theorem}
\textbf{Proof:} We define the datasets $E$  and $E^{\prime}$ as the set $\{e(t_i)\}_{i=1}^n$, and $\{e(t_i)^{\prime}\}_{i=1}^n$ where $e$ and $e^{\prime}$ correspond to the tracking error of the system (\ref{eq:plant}) obtained using the state feedback funnel controller with funnel boundary $\psi_i$ and $\psi_i^{\prime}$. From (\ref{eq:e_ineq_state}), we have
\begin{equation}
    -\psi(t_i) < e(t_i) <\psi(t_i) , \ \ \text{and} \ \ -\psi^{\prime}(t_i) < e^{\prime}(t_i) <\psi^{\prime}(t_i)
    \label{eq:f_ineq}
\end{equation}
for $i=1,\ldots,n$. From the above inequalities, we can conclude that the datasets $E$ and $E^{\prime}$ are in the interior of the datasets $\Psi$ and $\Psi^{\prime}$. Using (\ref{eq:f_ineq}) 
$$|e(t_i)-e^{\prime}(t_i)| < |\psi(t_i)-\psi^{\prime}(t_i)|, \ \ 1\leq i \leq n $$
Using (\ref{eq:adj}) we have
\begin{equation}
    |e(t_i)-e^{\prime}(t_i)|   < \delta \psi, \ \ 1\leq i \leq n
\end{equation}
From Theorem 5, the funnel boundary is $(\epsilon,\delta)$ differentially private for all $t \in [\bar{T},T_f]$. For $t_i \in[ \bar{T}, T_f]$, we define datasets $\bar{E}$, $\bar{E}^{\prime}$ as the set $\{e(t_i)\}_{i=n_1}^{n_2}$, and $\{e(t_i)^{\prime}\}_{i=n_1}^{n_2}$ where $n_1$ and $n_2$ can be determined from $\bar{T}$, and $T_f$. Using the above inequalities and (\ref{eq:adj_new}) we have
\begin{equation}
    |e(t_i)-e^{\prime}(t_i)|   <  \delta \psi^{\prime}, \ \ n_1\leq i \leq n_2
\end{equation}
Moreover,
\begin{equation}
    -\psi(t_i) < e(t_i) <\psi(t_i) , \ \ \text{and} \ \ -\psi^{\prime}(t_i) < e^{\prime}(t_i) <\psi^{\prime}(t_i)
    \label{eq:f_ineq_new}
\end{equation}
for $i=n_1,\ldots,n_2$. Therefore, we can conclude that the datasets $\bar{E}$, and $\bar{E}^{\prime}$ are in the interior of the dataset $
\bar{\Psi}$ and $\bar{\Psi}^{\prime}$ for all $t \in [\bar{T},T_f]$ and satisfies the adjacency relationship (\ref{eq:adj_new}). Therefore, the following holds \cite{le2013differentially}
\begin{align}
    \nonumber \mathbb{P}\left[\mathcal{M}\left(e\right) \in S\right] \leq \exp^{\epsilon} \mathbb{P}\left[\mathcal{M}\left(\psi\right) \in S\right]+\delta \text { for all } S \in \mathbb{R} .
\end{align}
\begin{align}
    \nonumber \mathbb{P}\left[\mathcal{M}\left(e\right) \in S\right] \leq \exp^{\epsilon} \mathbb{P}\left[\mathcal{M}\left(\psi^{\prime}\right) \in S\right]+\delta \text { for all } S \in \mathbb{R} .
\end{align}
\begin{align}
    \nonumber \mathbb{P}\left[\mathcal{M}\left(e\right) \in S\right] \leq \exp^{\epsilon} \mathbb{P}\left[\mathcal{M}\left(e^{\prime}\right) \in S\right]+\delta \text { for all } S \in \mathbb{R} .
\end{align}
Using the above we conclude differential privacy of the tracking error.

Next we prove the tracking error differential privacy for the output feedback funnel controller.
\begin{theorem}
\textit{
Let Theorem 2 and 5 hold and the state feedback funnel controller be defined by (\ref{eq:outputfeedback}) and the mechanism $\mathcal{M}:Q(d)+ y(t)$ where $y$ is generated from the probability distribution (\ref{eq:bounded_Gauss}) and the OU type process (\ref{eq:OU}) makes the funnel boundary $(\epsilon,\delta)$. Then, the tracking error is $(\epsilon,\delta)$ differentially private for all $t \in [\bar{T},T_f]$ for some finite time $T_f>\bar{T}>0$.}
\end{theorem}
\textbf{Proof:} The proof can be done by repeating the steps of Theorem 6 and using the relation (\ref{eq;e_ineq_out}).


\section{Simulation Results}
\subsection{Numerical Simulations for the OU type process}
\begin{figure}[h]
 \includegraphics[width=\linewidth]{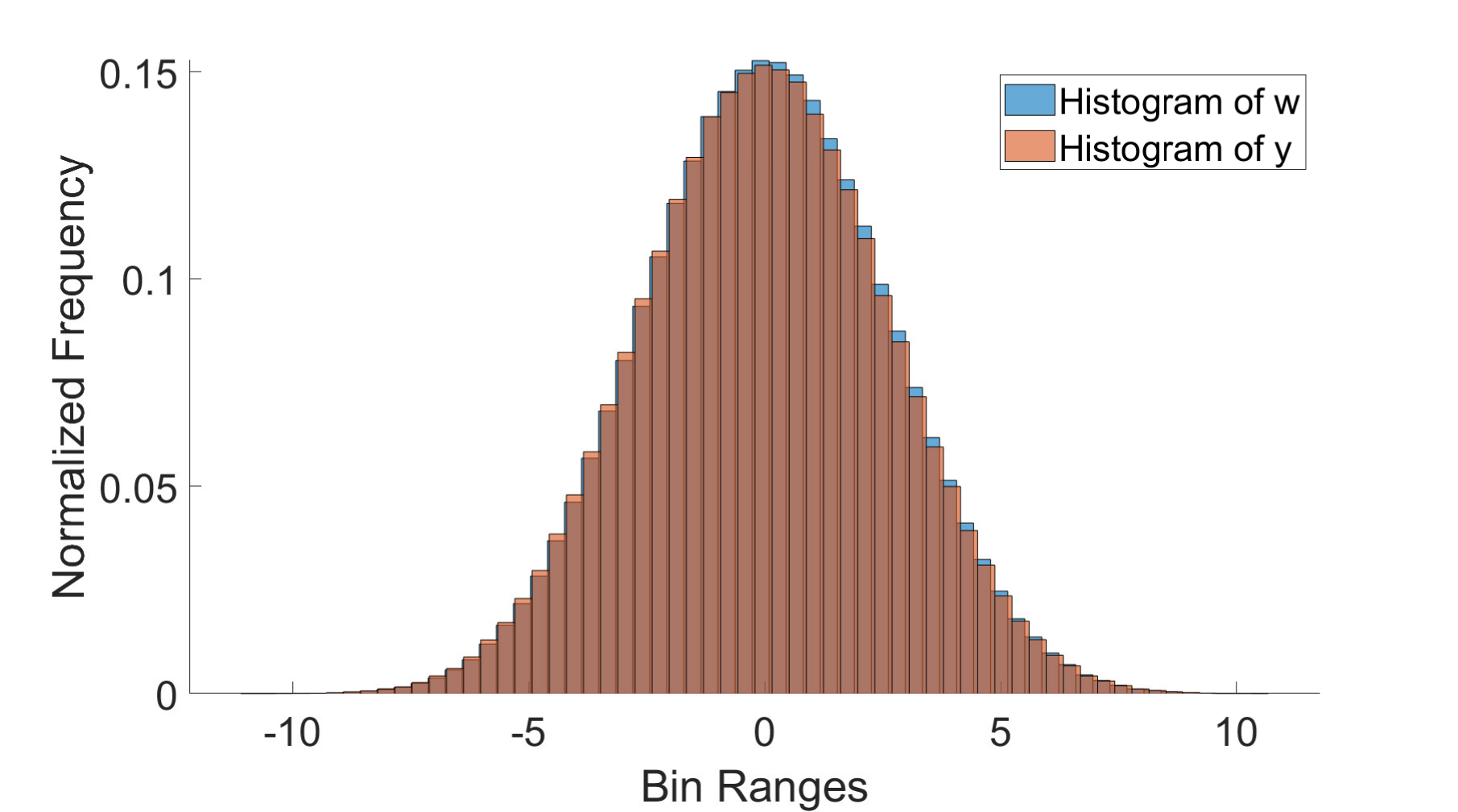}
\caption{ Comparison of the histogram of $y$ and $w$ }
\label{Fig:hist}
\end{figure}

We simulate a discrete OU type process of the form:
\begin{subequations}
\begin{align}
\begin{split}
y(k+1)= a_y y(k)+b_y w(k)
\end{split}\\
\begin{split}
w(k+1)= a_w w(k)+ b_wv( k)
\end{split}
\end{align}
\label{eq:OU_discrete}
\end{subequations}
where $v$ is sampled from a truncated Gaussian distribution. The system is simulated with parameters $a_y=0.01,b_y=1,a_w=0.9,b_w=1$, with initial conditions $y(0)=w(0)=0$. Table \ref{table:comp} shows that the variance and mean of $y$ and $w$ are very close to each other since from Theorem 3 we have $y=w+O(\theta)$. Fig. \ref{Fig:hist} shows the histogram of $y$ and $w$ for the last scenario in Table \ref{table:comp}. It can be seen from the figure that the two histograms overlap each other with a very small deviation.
\begin{table}
\caption{Comparison of mean and variance of $w$ and $y$ of the OU type process}
\small
\centering
\begin{tabular}{cccccc}
\hline
 $(\mu,\sigma)$ & $(\alpha,\beta)$ & Var($y$) & Var($w$) & $E[y]$ & $E[w]$ \\ \hline
  (0,1) & (-0.5,0.5)          &  0.4323   &   0.4246 & -5e-4   &   -5e-4\\
 (0,1) & (-1,1)          & 1.5615   &  1.5335 & -0.0011   &   -0.0011\\
(0,3) & (-1.5,1.5)    &  3.8909 &      3.8211&   -0.0016 & -0.0016 \\
(0,3) & (-2,2)         &  6.7378 & 6.6169  &    -0.0022 &  -0.0022 \\ \hline
\end{tabular}
\label{table:comp}
\end{table}

\subsection{Tracking Error $(\epsilon,\delta)$ Privacy Results}
We consider a second-order nonlinear system
\begin{subequations}
\begin{align}
\begin{split}
\dot{\xi}_1 =\xi_2
\end{split}\\
\begin{split}
\dot{\xi}_2=-\xi_{1}+\xi_{1}^3+u
\end{split}
\end{align}
\label{eq:nonlinear_system}
\end{subequations}
The reference signal is generated from the output of the following exosystem \cite{chowdhury2020practical}
\begin{subequations}
\begin{align}
\begin{split}
\dot{\xi}_{r1}=\xi_{r1}
\end{split}\\
\begin{split}
 \dot{\xi}_{r2}=2(1-\xi_{r1}^2)\xi_{r2}-\xi_{r1}
\end{split}
\end{align}
\label{eq:nonlinear_system_reference}
\end{subequations}
where $r(t)=\xi_{r1}$. The performance funnel is chosen as
$$\psi(t)=(2\pi-\psi_{ss})e^{-t/2}+\psi_{ss}+y(t)$$
where $\psi_{ss}$ is the steady-state bound and $y(t)$ is the output of the Ornstein Uhlenbeck process. We transform the system into the error coordinates and choose the virtual  output as,
\begin{equation}
s=\omega_1+k_2\omega_2
\end{equation}
where $\omega_1=\xi_1-r$, $\omega_2=\varrho(\xi_2-\dot{r})$ and $e=\omega_1$. 
The output feedback funnel controller is given by
    \begin{subequations}
    \begin{align}
    \begin{split}
    u&=\frac{1}{\varrho^2 k_2}\left[\frac{1}{k_2}\left(\hat{s}_s-\hat{\omega}_{1s}\right)+\varrho \hat{v}_{fs} \right]
    \end{split}\\
    \begin{split}
     \hat{k}(t)&=\frac{1}{\psi(t)-|\hat{s}_s(t)|},\ \ \hat{k}_s(t)=\bar{M}_k \text{sat}\left(\dfrac{\hat{k}}{\bar{M}_k}\right) 
    \end{split}\\
    \hat{v}_{fs}&=-\hat{k}_s(t)\hat{s}_s(t)
    \end{align}
    \label{eq:controller sim}
    \end{subequations}
where $M_k=5$ and the estimates are given by the high-gain observer
    \begin{subequations}
    \begin{align}
    \begin{split}
    \dot{\hat{e}}_1&=\hat{e}_2+\frac{\gamma_1}{\varsigma}(e-\hat{e}_1)
    \end{split}\\
    \begin{split}
     \dot{\hat{e}}_2&=\frac{\gamma_2}{\varsigma^2}(e-\hat{e}_1)
    \end{split}
    \end{align}
    \label{eq:observer_sim}
    \end{subequations}    
The estimates $\hat{\omega}_1=\hat{e}_1$ and $\hat{\omega}_2=\varrho \hat{e}_2$ are saturated with the saturation levels $\pm 1$ and $\pm 3$. The saturation levels are chosen from simulations to see
the maximal values that the state trajectories would take when using the state feedback controller. The simulation is carried out with $\xi_1(0)=2, \xi_2(0)=0$, $\xi_{r1}(0)=1, \xi_{r2}(0)=1$, $k_2=7.5$, $\varrho=0.01$, $\varsigma=0.001$, $\gamma_1=2$, $\gamma_2=1$, $\psi_{ss}=1.3$, $\alpha=-0.9$, $\beta=0.9$, $\delta
\psi =0.5$, $\mu=0$, $\sigma=1$. 

\begin{figure}[h]
 \includegraphics[width=\linewidth]{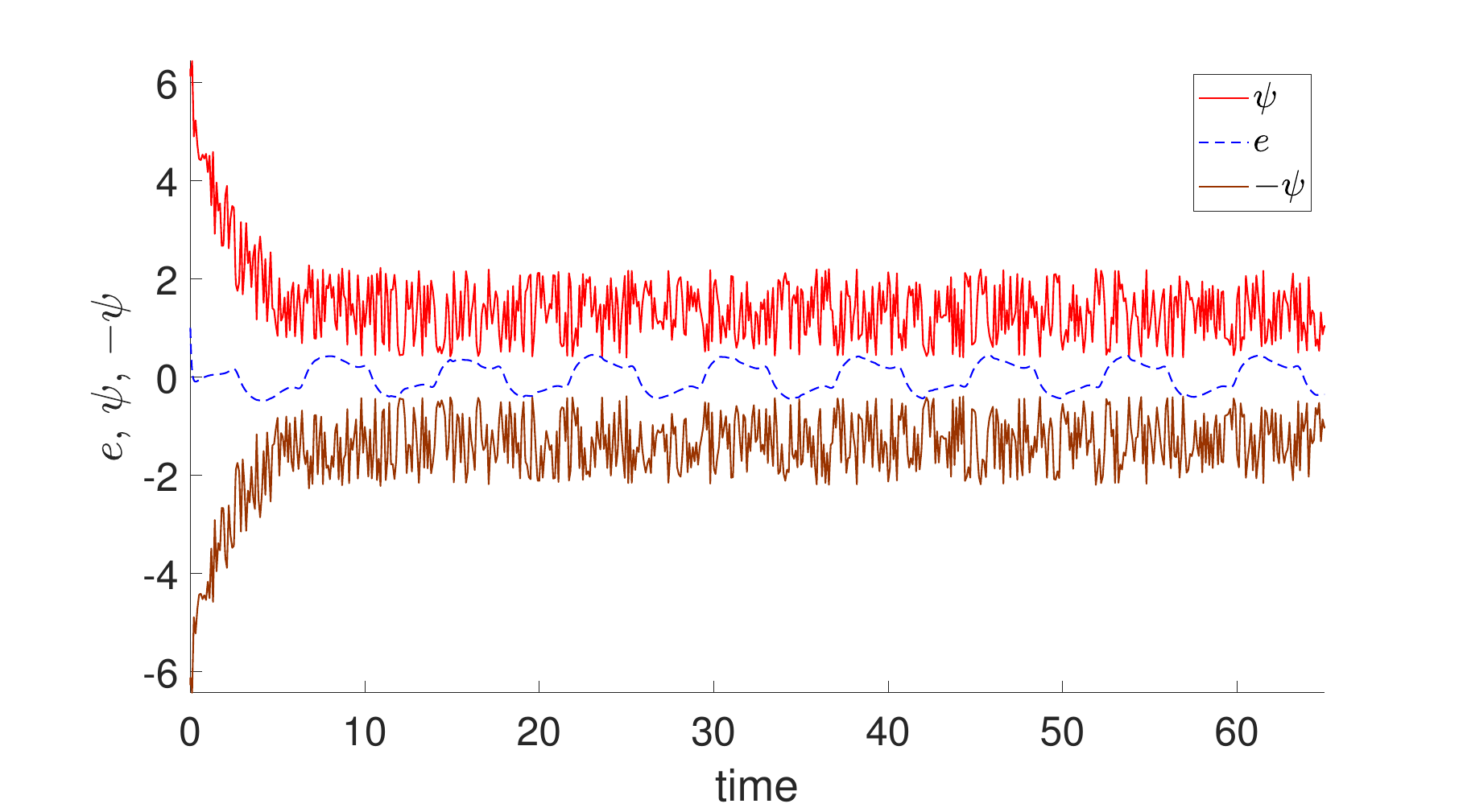}
\caption{Tracking error evolution inside performance funnel}
\label{Fig:error_evolution}
\end{figure}

\begin{figure}[h]
 \includegraphics[width=\linewidth]{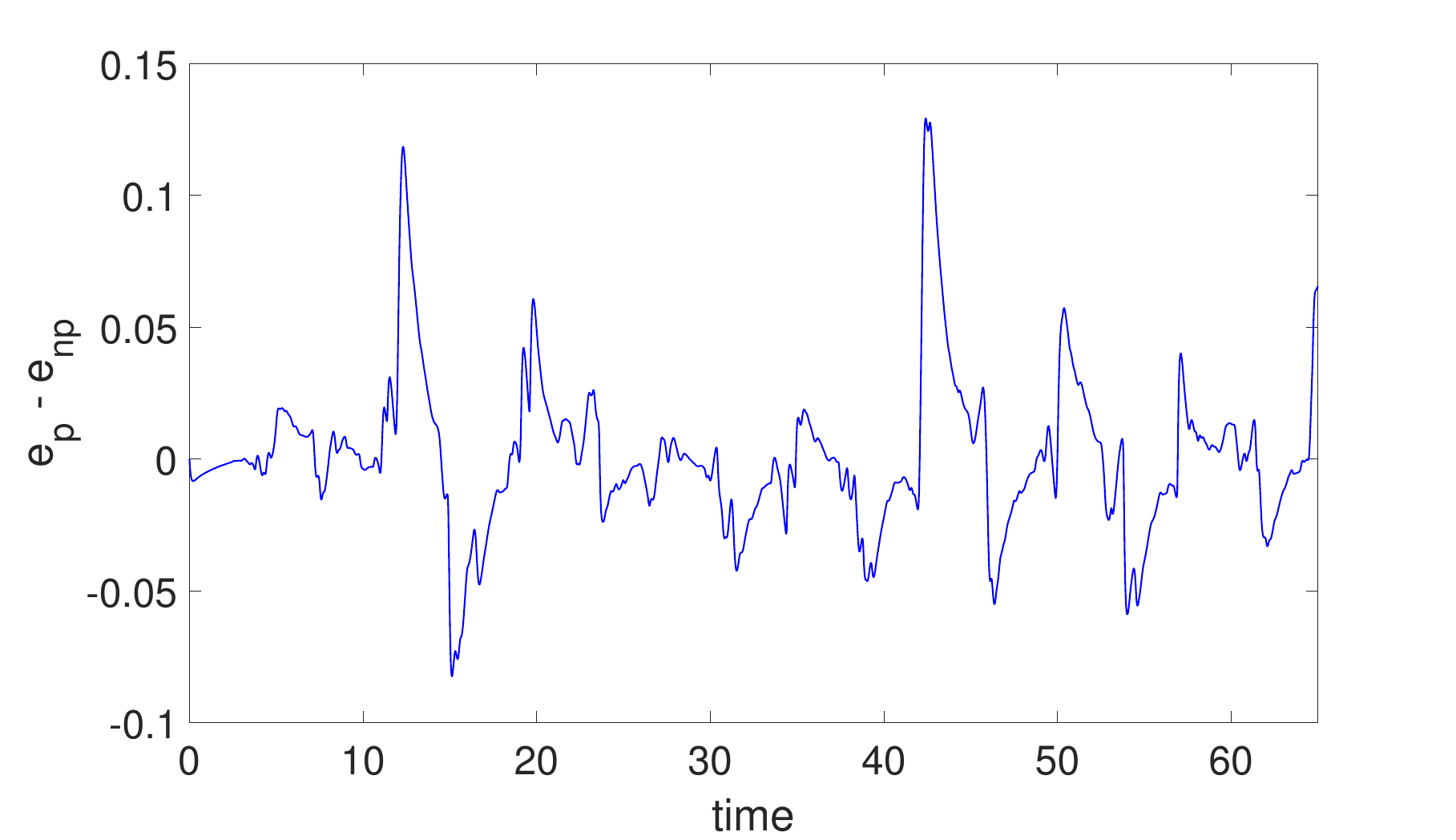}
\caption{Difference in tracking error in the presence privacy signal}
\label{Fig:tracking_error_diff}
\end{figure}

\begin{figure}[h]
 \includegraphics[width=\linewidth]{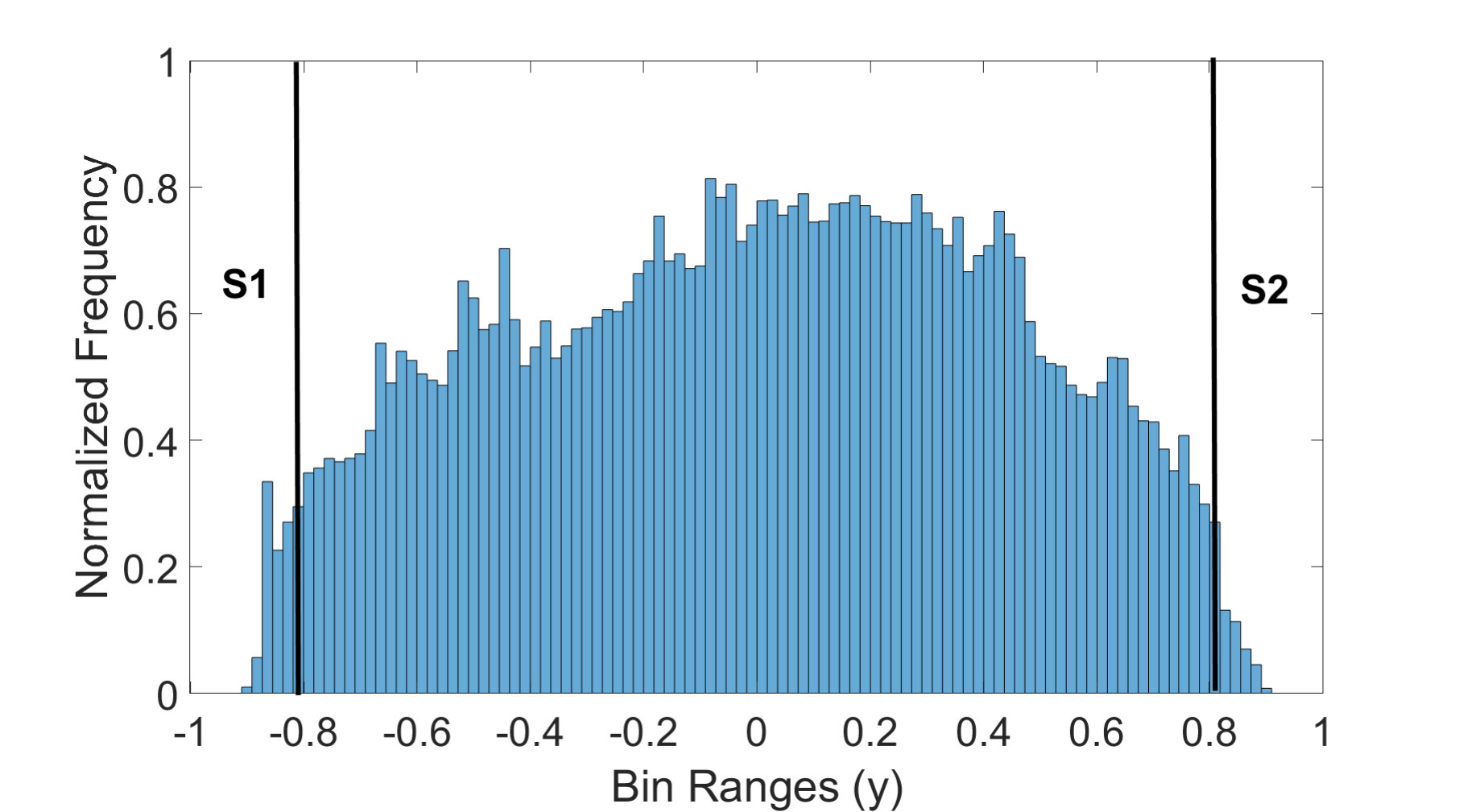}
\caption{ Histogram of $y$ with area divisions for $\epsilon$, $\delta$ calculation }
\label{Fig:hist_y_e_delta}
\end{figure}

Fig. \ref{Fig:error_evolution} shows the evolution of $e$ within the performance funnel under the output feedback controller (\ref{eq:controller sim}). Fig. \ref{Fig:tracking_error_diff} shows the difference between the tracking errors in the presence and absence of privacy signals. Fig. \ref{Fig:hist_y_e_delta} shows the histogram of the output of the OU type process $y$. In the figure, for the choice of $M=0.8$, the areas $S1$, $S2$ are shown for the calculation of $\epsilon$, and $\delta$ which is calculated as
$$\epsilon \leq   1.0001, \ \ \text{and} \ \ \delta \leq 0.0397 $$

\section{Conclusion}
In this paper, we presented a new framework for introducing differential privacy in the tracking error of nonlinear systems. The initial funnel control problem is formulated for nonlinear systems with arbitrary relative degrees using high-gain observers using the idea of the virtual output. We then make the performance funnel differentially private by adding a continuous bounded noise which is the output of an Ornstein-Uhlenebck type process. We provide bounds of $\epsilon$ and $\delta$ using the results of \cite{he2020differential} and show that the tracking error is differentially private using the differential privacy of the performance funnel. 

\bibliographystyle{IEEEtran}
\bibliography{Refs}


\end{document}